# Authentication and Access Control in 5G Device-to-Device Communication

Jithu Geevargheese Panicker
*Faculty of Information Technology*
*Monash University*
Melbourne, Australia
jgee0004@student.monash.edu

Ahmad Salehi S.
*Faculty of Information Technology*
*Monash University*
Melbourne, Australia
ahmad.salehishahraki@monash.edu

Carsten Rudolph
*Faculty of Information Technology*
*Monash University*
Melbourne, Australia
carsten.rudolph@monash.edu

*Abstract*—Device-to-device (D2D) communication is one of the most recent advancements in wireless communication technology. It was introduced in cellular communication technology by the $3^{rd}$ Generation Partnership Project (3GPP) to lay a foundation for the evolving 5G architecture. It has now emerged as a promising technology for proximate devices. It enables proximate devices to communicate directly without the involvement of a third party network infrastructure. Researchers are analysing various methods to facilitate the smooth integration of D2D communication technology into the existing network system architecture. This paper lists all the different possible modes of operation in D2D communication based on the varying use-case scenarios and highlights the security and privacy requirements for D2D communication. Some of the recent authentication proposals for D2D communication technology are further reviewed, and their security and privacy capabilities are analysed. Apart from authentication, we also reviewed some recent proposals of access control in D2D and highlighted the security issues addressed. We then identified the open issues that prevail in implementing D2D technology in a real-world scenario for future researchers, emphasising the existing authentication and access control techniques in D2D communication.

## I. INTRODUCTION

The world is currently witnessing the transformation into the fifth generation of wireless mobile communication technology. Device-to-device (D2D) communication is one of the significant features in Release 12 of the $3^{rd}$ Generation Partnership Project (3GPP) [1]. The mobile broadband standard community, 3GPP, introduced D2D to lay a foundation for the evolving 5G architecture and facilitate off-grid communication. Traditional mobile cellular communication technology relies on a network infrastructure system comprising base stations (BSs) and a core network (CN) to communicate between the devices. Data transferred from source to destination is always routed via the cellular network infrastructure, even if they are close to each other. D2D enables multiple devices to communicate without traversing via intermediate access points (APs) and BSs, thus reducing the CN dependency. However, an increase in several devices using the traditional cellular system deteriorates the system's reliability due to limited bandwidth capacity and options for scalability. Recent researches such as [2]–[4] demonstrates the importance of D2D communication for the present and the future.

Effective authentication of D2D devices that try to communicate with each other is critical. These devices generally use symmetric and asymmetric keys to initiate a session and encrypt data during transmission. Unlike regular computers, IoT devices are resource-constrained due to their strip-down sizes and processing power. Therefore, authenticated devices are granted varying levels of access to control the flow of data between communicating devices. These levels are established with the help of access control policies set by the owner or manufacturer of the device that is involved in the communication [5].

Security and usability will always remain the biggest concerns for any upcoming technology [6]. Current works [7]–[9] demonstrates various methods to integrate D2D communication system architecture into the existing network system architecture. This paper primarily tries to analyse the recent authentication and access control proposals for D2D. After analysing various research works, we intend to highlight the major open issues in D2D communication and its future research scope. Throughout the paper, we mention D2D for D2D in 5G since all the methods discussed in the following sections can be extended to the 5G architecture. Our main contributions through this paper are as follows:

i. Identify the different modes of operation for D2D communication.
ii. Conduct a review of the recent authentication proposals in D2D communication and analyse their security and privacy capabilities.
iii. Conduct a review of the recent access control proposals in D2D communication and highlight the security issues addressed.
iv. Identify the open issues that stand as a roadblock for implementing D2D communication to provide future research directions for the research community.

The remaining sections in this paper have been organised as follows. In section II we present a background on D2D and other relevant wireless communication technologies. In section III and IV, we discuss the security and privacy requirements of D2D communication and the recent authentication and access control techniques in D2D. The open issues prevalent in D2D communication are discussed in section V for future researchers. Finally, we conclude the paper in section VI.

## II. BACKGROUND

This section reviews D2D, cellular, and other relevant wireless communication technologies that can facilitate wireless transmission between proximate devices.

*A. D2D in 5G and other relevant wireless communication technologies*

D2D communication technology enables multiple devices to communicate without trailing via the AP, BS and the CN [10]. In traditional cellular technology, the exchange of data between the user equipment (UE) and the BS requires a specific bandwidth from within the licensed spectrum to establish communication between the two. UE devices are devices used by the end-users to establish communication.

Any UE located outside the network coverage area cannot establish communication with other devices via cellular technology. However, they could communicate via short-range D2D communication technologies such as Bluetooth, Radio-frequency identification (RFID), and infrared, where data can be directly transferred. The significant challenges in a cellular network would be the over-utilisation of the bandwidth and higher energy consumption due to the limited spectrum [2]. Previous generations of cellular technology ranging from 1G to 4G were able to provide data speeds based on the use-case scenarios of the specific period [2]. However, none of them could address the significant challenge of cellular networks mentioned above due to the devices' operating frequency range limitation. Nevertheless, with the rapid increase in the number of devices, the primary issue remains in accommodating all these devices into the existing cellular communication systems [11]. The 5G infrastructure systems can address the significant issue of helping proximate devices to communicate with each other, even in the absence of network connectivity via the incorporated D2D communication technology.

*1) D2D Communication categories based on application and use-case scenarios:* To date, there are various applications and use case scenarios that have evolved based on D2D communication. For instance, IoT, pervasive social networking [13] and content sharing [14]. Based on the network infrastructure, applications and use case scenarios, D2D communication can be classified into three wide categories as illustrated in Fig. 1 [12], [7].

In-coverage mode: In this mode, all the UE's are located within the network's coverage area, as shown in Fig. 1a. These devices are fully controlled by the cellular network infrastructure comprising of the AP, BS and CN. The CN is responsible for the authentication of different UE's, resource allocation, connection establishment, and security and privacy management. This mode of D2D shares the licensed frequency spectrum with the traditional cellular network. The licensed spectrum establishes a connection and allocates the resources between the UE and the BS. Once a secure connection is established between these devices with the help of a CN, these devices can further communicate with each other leveraging the advantages of D2D communication [7].

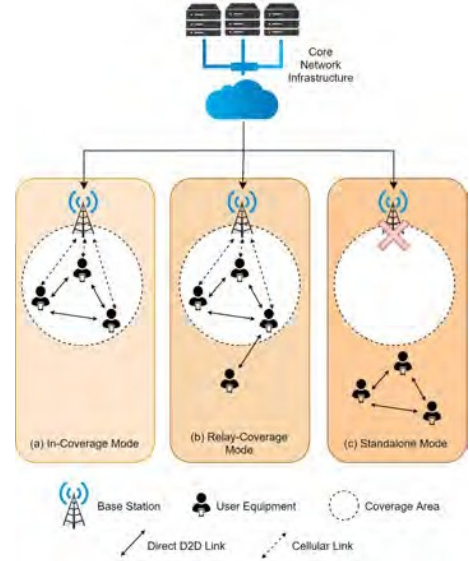

Fig. 1. D2D Communication categories based on the application and use-case scenarios: a) In-coverage mode, b) Relay-coverage mode, c) Standalone mode [12]

Relay-coverage mode: There may exist some UE's located beyond the BS's coverage area. These devices cannot access the BS or experience low connection quality while enabling communication between them. In this mode, D2D communication can enable the UE's beyond the coverage area of the BS to partially establish a connection with the BS via another UE that is present within the coverage area of the BS, as shown in Fig. 1b. The intermediate UE will act as a link between the remote UE and the BS. The CN does not directly control the device but can provide partial assistance in resource allocation, establish a connection, and ensure security management. In such scenarios, D2D communication can enable communication between the UE devices using the licensed or unlicensed spectrum [12].

Standalone mode: In this mode, the UE's can communicate with each other if the devices are close to each other but located far away from the BS's coverage area. The UE's can directly communicate with each other. A typical use case scenario for this mode would be emergency services or disaster relief when the communication systems are down. [15] demonstrated the importance of the standalone mode of D2D communication and how they become essential for emergency communication systems.

*2) Cellular Technologies:* The experiments conducted by Heinrich Rudolph Hertz can be considered the cornerstone of wireless communication technology. The second half of the twentieth century marked the dawn of wireless communication technology, pioneered by the technical activities conducted at Bell Labs. The evolution of cellular technology witnessed four generations of telecommunication technologies. Currently, we are experiencing transformation into the fifth generation of wireless telecommunication technology.

The fifth generation of the telecommunication standard was introduced in late 2018 as 4G's successor. The major companies in the cellular networks are currently focusing on the roll-out of 5G networks due to their advantages and enormous applications. It was proposed as a futuristic solution for applications that require a high data rate. Some of the underlying features [16] of 5G are:

i. 5G uses millimetre waves with an operating frequency of 30GHz - 300GHz. The wavelength varies from 1mm - 10mm and is hence known as millimetre waves.
ii. Small Cells are base stations that are small and portable. They are installed every 250 meters to provide a dense network for connectivity.
iii. Massive Multiple-Input Multiple-Output (MIMO) supports about 100 antenna ports which increase the capacity of the network.
iv. Beam-forming technology can allow the waves from these antennas to propagate in a specific direction towards the receiver to avoid interference and enhance the signal strength.
v. 5G supports Full Duplex connection where it allows transmission and reception of data simultaneously within the same frequency.

*3) Other Wireless Technologies:*

*a) Short Range Wireless Transmission Techniques:* Some of the most used short-range wireless communication techniques in the application of IoT data transfer between resource-constrained devices are:

i. Wireless Fidelity (Wi-Fi): It is a short-range wireless communication technology that uses the IEEE 802.11 standard [17]. It enables the implementation of Wireless Local Area Networks (WLAN). It can use the unlicensed radio bands, the 2.4GHz band and the 5GHz band for communication. WPA3 [18] is the latest security standard that is advised for Wi-Fi communication.
ii. Bluetooth: This short-range wireless communication technology uses the IEEE 802.15.1 standard for exchanging data between devices. It enables the implementation of Wireless Personal Area Networks (WPAN) due to its range limitations [19].
iii. ZigBee: ZigBee uses the IEEE 802.15.4 standard for enabling resource-constrained devices to communicate over a network and form a WPAN [19].
iv. Radio Frequency Identification (RFID): RFID is yet another short-range wireless technique used to identify, collect, and track information from an object (animals, humans). RFID readers are used to reading and write data from RFID tags [19].

*b) Long Range Wireless Transmission Techniques:* Long-range wireless communication technologies [20] provides a kilometre-wide range for data transfer between devices compared to traditional short-range wireless communication. Since the data is transmitted at longer distances, the data rate is reduced. Hence, they use the lower frequency band for communication. Advancement in these transmission techniques resulted in the Low-Power Wide-Area networks (LPWAN). Industries and Academicians show much interest in these technologies due to their varying applications in D2D communication. SigFox and LoRa are some of the pioneers in long-range wireless transmission. These technologies use the unlicensed band for communication [21].

i SigFox: It is a long-range wireless transmission technique that provides long-range communication at low frequencies. It provides an end-to-end IoT connectivity solution. Proprietary base stations are employed with IP-based network connection. It employs the ultra-narrow band, thus provides low power consumption, higher receiver sensitivity, and low transfer rates at 100 bits per second [21].
ii LoRa (Long Range): It is a long-range wireless communication technique operating at the lower frequency range by modulating in a sub-GHz ISM band. The data rate varies from 300 bits per second to 50 kilo-bits per second [21].

## III. SECURITY AND PRIVACY REQUIREMENTS IN D2D COMMUNICATION

The demand for broadband wireless communication services is increasing day-by-day [22]. Nowadays, it is easier to manufacture devices that leverage these wireless technologies at minimal costs. However, most developers and manufacturers nowadays focus on implementing different systems and compromising security issues and the threats that can affect these systems' performance. D2D communication establishes a direct connection between the devices. Hence the devices are prone to many security threats such as data fabrication, user privacy violation, modification [23]. This section introduces some of the security and privacy requirements that are to be considered in D2D communication.

According to [12], authentication techniques for the in-coverage mode, which illustrates the communication between the CN and the UE via BS, have been thoroughly studied in the existing works [1]. [12] demonstrates the need for further research in the new entities and references such as the device authentication in standalone mode. Failing to authenticate these devices properly can lead to attacks related to the identity of a device, like masquerading and impersonation attacks. Ordinary devices with a SIM card can use their respective authentication and key exchange (AKA) [24] protocol via an AP. Once they reach out of the AP coverage area, they have to rely on other authentication modes. Since the data is sent wirelessly, anyone can access the data that is transmitted. It can lead to attacks such as tampering and eavesdropping attacks [23].

Table I provides a security and privacy analysis of the D2D communication authentication techniques discussed in section IV-A. We need to understand the relevance of these security and privacy requirements before analysing the authentication techniques with these requirements. The critical security and

privacy requirements for the practical implementation of D2D communication are mentioned below:

*Authentication and Authorisation:* Authentication is an essential step in identifying a device by verifying some key or a secret in possession of the device [7]. Once a UE is authenticated, the UE gets some predefined access to the UE's resources and data exchanged between the devices. Therefore, it is essential to satisfy this security requirement to ensure secure communication between UE devices in D2D.

*Non-Repudiation:* Non-repudiation ensures that the action of a UE sending a message cannot be denied, enabling the UE's to trust the data that is exchanged. It will also hold as legal proof in proving the authorship or the validity of the message [12].

*Confidentiality and Integrity:* Confidentiality ensures that only the UE devices that are authenticated to access the transmitted information can access them. Data should be encrypted before transmission. Integrity ensures that the receiver receives an exact copy of the message that is sent by the sender without any modifications [7].

*Privacy:* An attack on privacy can be viewed as a significant threat for D2D devices. Malicious devices try to access the parameters of a device. These device parameters include location, user information. Once accessed, these parameters can be critical at times and reveal the information about the subject required by the malicious device. It is not safe to reveal critical information to UE devices that have never met before. Furthermore, it can open doors to new problems such as location spoofing and eavesdropping attacks [23].

*Availability and Dependability:* When a UE is not available for communication even though it is available, then the availability and dependability [7] of the system is compromised. The security requirement's objective is to make sure the UE's can communicate even during attacks. During a Denial of Service (D0S) / Distributed Denial of Service (DDoS) or jamming attacks [23], the availability and dependability of the devices are compromised.

*Forward and Backward secrecy:* Forward secrecy ensures that a UE will not have access to any further data exchanged in a D2D communication group once it leaves the group. Backward secrecy ensures that a UE does not have access to the previous messages of a D2D communication group when it joins the group [12].

## IV. Taxonomy of Authentication and Access Control Techniques for D2D Communication in 5G

Authentication refers to how the identity of an entity that wishes to get access to a system is verified. Access Control refers to the policies set by the system for a particular entity to access its resources. Authentication and Access Control are the essential features of a system that can ensure adequate security [29]–[31].

### A. Authentication Techniques for D2D Communication

This section discusses some of the recent authentication techniques used to authenticate UE devices in D2D communication.

M. Wang and Z. Yan proposed two AKA protocols for D2D communication; PPAKA-HMAC and PPAKA-IBS [25]. These protocols enabled D2D group communication securely and anonymously. PPAKA-HMAC helps in establishing a secure group session. The protocol is identified to be secure against attackers and resource-constrained applications. PPAKA-IBS is an extension of PPAKA-HMAC and is capable of preventing internal attacks. Security and experimental tests were further performed over these protocols to test the effectiveness of the protocol.

In [14], A. Zhang et al. proposed a secure data sharing protocol with the approach of cryptography, which uses both public-key-based signature and symmetric encryption to ensure the security needs of the system. Data transmitted is signed (public-key encryption) by the data provider via BS and further re-signed by other transmitters (BS) to ensure transmission non-repudiation and verify the UE. The receiver must send a key-hint request to decrypt received data from BS to achieve reception non-repudiation. The receiver can verify the integrity of the data received by verifying the signature. Symmetric encryption is used to enable integrity and confidentiality by end-to-end encryption. The device authentication occurs between the UE and the BS by verifying the signature at the receiver UE. The UE devices cannot be authenticated in the absence of the data provider.

RF fingerprint identification technology was used by Z. Zhang et al. in [26] unlike traditional cryptographic mechanisms and outdated security protocols to authenticate D2D devices. The device recognition rate has always been accurate when the signal-to-noise ratio (SNR) is over 8dB. The RF fingerprint was generated using the Hilbert transform and the PCA method to identify wireless devices. The proposed method helps in managing trust between D2D devices apart from improving security.

An authentication method that uses a Secret Unknown Ciphers embedded machine learning model to deploy users keystroke dynamics with accelerometer biometrics for creating user identification profiles was proposed in [27]. For two devices to establish a connection, both the devices request identification from the TA. The TA uses the information of the devices' biometric identity and mobile identity to identify the requests. Once the identities of the devices are matched, they can communicate with each other.

G. Lopes et al. proposed a mutual authentication protocol for IoT-enabled mobile-health (m-health) systems that are capable of communicating via D2D in [9]. There is a considerable amount of sensitive data exchanged between devices and are required to meet the privacy expectations of the patients using these systems. The use of symmetric cryptography over asymmetric provides better performance for resource-constrained devices and reduces the system's cost without compromising on-device security.

D. Yang et al. tries to correlate the direct wireless communication strategy of D2D with the blockchain-based communication over a peer-to-peer (P2P) network in [28]. The paper proposes a consensus algorithm that can facilitate D2D communication and a selective blockchain system (SBC). The proposed system can identify the security of the D2D network and the security resources of the participating devices that are joining the blockchain consensus. The proposed SBC includes the off-chain and on-chain scenarios that are present in D2D communication.

TABLE I
SECURITY AND PRIVACY CAPABILITIES OF SOME OF THE RECENT RESEARCH WORKS ON AUTHENTICATION IN D2D COMMUNICATION

| Ref. | Year | Technique Used | Security and Privacy Requirements | | | | | |
|------|------|----------------|------|------|------|------|------|------|
| | | | A/A | NR | C/I | Pr | A/D | F/B |
| [14] | 2016 | Secure Data Sharing Protocol | Y/Y | Y | Y/Y | Conditional | Y | -/- |
| [25] | 2018 | PPAKA-HAMC and PPAKA-IBS | Y/Y | Y | Y | Y | - | Y/Y |
| [26] | 2018 | RF Fingerprint Identification | Y/Y | - | -/- | - | - | -/- |
| [27] | 2019 | Secret Unknown Cipher, Machine Learning | Y/Y | Y | Y/Y | Y | - | -/- |
| [9]  | 2020 | Mutual authentication protocol for m-health systems | Y/Y | Y | Y/Y | Y | Y | Y/Y |
| [28] | 2021 | Blockchain over P2P | Y/Y | Y | Y/Y | Y | Y | -/- |

[32] discusses some of the security aspects and challenges in D2D communication. The paper proposes a secure and efficient protocol for the transmission of data in D2D communication. It uses the Diffie-Hellman key agreement and commitment schemes to authenticate the devices and enable communication. It enables a device to establish a shared secret key for communicating with devices without prior knowledge. Even though these devices are end-to-end encrypted, there is still no possibility to identify the identity of the device claimed by the same is genuine.

Similar to [32], [33] presents two secure channel authenticated key exchange protocols for enabling end-to-end security. It provides group anonymity to the UE's by hiding the identity and the information of the group after authentication [34] while communicating between devices. These devices store the identity of devices of the same group to identify a specific device and block communication from the other devices while authenticating. It provides group anonymity in all the modes of D2D communication.

Authentication in short-range wireless communication in D2D communication does not provide a satisfactory solution. Technologies like Bluetooth requires manual pairing, and Wi-Fi needs access to an AP that is user-defined. [35] discusses the shortcomings of short-range wireless communication technologies such as Wi-Fi and Bluetooth. Wi-Fi and Bluetooth require manual pairing before establishing a connection between devices when connecting for the first time. These devices cannot connect and share data without human intervention, acting as a roadblock for autonomous and proximity-based services dynamically.

In traditional cellular communication technology, UE devices are equipped with a SIM card which is used to authenticate the UE device [24]. Apart from the authentication techniques discussed previously, 5G - Extensible Authentication Protocol-Transport Layer Security (5G-EAP-TLS) protocol was introduced in 5G, which allows devices to communicate in private networks and IoT networks via cellular communication technology [8].

A security and privacy requirement analysis of some of the various recent authentication techniques in D2D has been illustrated in Table 1. The requirements not relevant or not discussed in the respective article has been left blank. The abbreviations for the headings used in Table 1 are as follows: A/A - Authentication/Authorisation; NR - Non-Repudiation; C/I - Confidentiality/Integrity; Pr - Privacy; A/D - Availability/Dependability; F/B - Forward Secrecy/Backward Secrecy

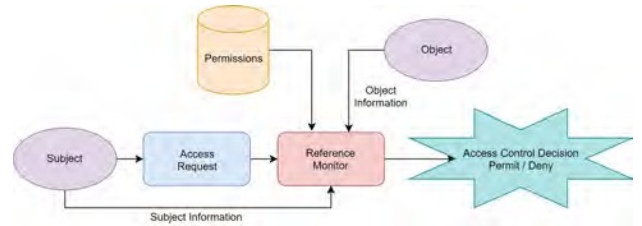

Fig. 2. Generic Access Control Architecture [42]

B. Access Control Techniques for D2D Communication

Once authenticated, access control is the next phase. Access control refers to the policies set by a particular system to ensure that the devices can access only the data and resources that they are allowed to access under varying circumstances. Fig. 2 shows a generic access control architecture in which a device requests access to protected data or resources to perform some action. The access control mechanism checks if the device is permitted to perform the action based on the access control policies. These policies are decided based on various factors, including the permissions, the device's role, the context and the relationship between the device and protected data or resource. Access control techniques can be divided into two types [43], namely, dynamic and static.

In D2D communication, once the UEs are authenticated, the access provided to the participating UEs to share and collect information from each other is controlled. There are two different aspects of research interests about access control as identified—access control on physical resources and access control on private information. In the following, we will discuss some of the recent access control techniques observed

TABLE II
SUMMARY OF THE SECURITY ISSUES AFFECTED AND TECHNIQUES USED IN OF SOME OF THE RECENT RESEARCH WORKS ON ACCESS CONTROL IN D2D COMMUNICATION

| Ref. | Year | Access control endpoint | Technique Used | Security issues addressed |
|------|------|------------------------|----------------|---------------------------|
| [36] | 2016 | Physical resources | Ad hoc On-Demand Distance Vector | - |
| [37] | 2018 | Private information | Makes use of a device whitelist mechanism | Trust management, Prevention from DoS and masquerading attacks, overhead reduction |
| [38] | 2018 | Private information | Uses 2 levels of trust and ABE | Trust level management, lightweight, Identity privacy, flexibility, scalability, confidentiality |
| [39] | 2018 | Private information | Robust and scalable data access control scheme | Fine-grained access control, Data Confidentiality, UEs Collusion Resistance, Data Integrity |
| [40] | 2018 | Physical resource | Multi-hop relaying between UE nodes | - |
| [41] | 2019 | Private information | Makes use of blockchain and smart contracts for authorisation | Immutable and transparent logging of data exchange, Prevention against DoS attacks |

under both the research aspects, emphasising personal information access control.

In [36], M. Murali and R. Srinivasan proposed the Ad hoc On-Demand Distance Vector (AODV) protocol. The system can transmit cached data between devices connected through a network of connected devices in a Mobile Ad-hoc Network. The group access control feature discussed in the paper can further be expanded into D2D communication to enable secure data access control. The model is similar to the standalone mode of D2D communication but yet a primitive one. The security analysis of the system was not performed in the same.

S. You et al. identified the threat of interaction between unknown devices and unnecessary calculations as the main challenges of D2D communication in [37]. The paper proposes a secure method of authenticating and providing access control using a device whitelist mechanism that focuses on trust management. The proposed system can prevent security issues such as personal information leaks, denial of service, and masquerading in D2D, which is standard in communication systems. It also provides a certain level of D2D authorisation, trust management and overhead reduction. The protocol can provide fine-grained access control with the help of parameters, but the device's true identity cannot be verified. Furthermore, the whitelist parameters, if known, can establish malicious connections if security policies are not vital for the initial connection phase.

In [40], D. Ebrahimi et al. tries to investigate the issues in the transmission of data between devices that can communicate via D2D links through multiple nodes. These multiple nodes would remove the network backhaul traffic reliance and use different possible D2D links to transfer the data to the destination device. The relaying of data through multiple hops can be entertained as long as the devices do not experience a delay in data transmission. The data communication is redirected to conventional BS for any D2D communication pair that experiences a transmission delay, as seen in a typical cellular communication network. The preset delay time frame ensures the quality of service of the approach from being compromised. The proposal's security analysis is yet to be completed and prone to security threats and attacks.

Z. Yan et al. identified the issues in secure D2D communication in D2D communication among mobile device and proposed a trust-based access control scheme for D2D mode communication in [38]. The scheme uses two levels of trust and applies Attribute-Based Encryption (ABE) to control access between the D2D communication link [44]. The two levels of trust are General Trust and Local Trust. General trust is issued by the CN, whereas the device analyses the local trust. A device can also incorporate both these trust model into the system. However, the local trust is not as efficient as the general trust. Hence, UE's cannot establish maximum security in the standalone mode of D2D. On the other hand, the proposed model works well with the In-Coverage and standalone mode.

Q. Li et al. try to address three issues that were identified in D2D communication access control in [39]. The three issues were data security, the privacy of identity, and the system's scalability. A data access control scheme is proposed in the system with multiple authorities and attributes. Every BS in the system was capable of generating intermediate attribute keys based on the scheme. The CN would further create private keys based on the intermediate attribute keys. The UE verification always requires a BS and CN, making it suitable for the In-Coverage mode of D2D communication. A single CN server only has to generate private keys for all the UEs in the system. The paper does not consider the standalone mode of D2D communication.

V.A. Siris et al. proposed two approaches for providing authorisation using blockchain and smart contracts in [41]. In the first approach, the blockchain records the hashes of the authorisation transactions related to payment links into the blockchain. The authorisation requests are accepted using a smart contract in the second approach. The cost analysis of the system is also provided in the paper. A local Ethereum based node with Go-Ethereum3 was used to implement the blockchain. The smart contract was written using Solidity in Remix-IDE. The most significant advantage of using blockchain is immutability and transparency. The policies can be set based on events in a smart contract. However, the client must always be connected to enable D2D communication.

The relay-coverage mode is used in the proposed method, and the approach does not support the standalone mode. This work opens a new prospect for providing authorisation using blockchain than the other access control techniques mentioned earlier. The security issues addressed and the techniques used for all the reviewed papers on recent access control techniques in D2D communication discussed above are highlighted in Table 2.

## V. Open Issues for Future Researchers

We are currently living in a world where billions of people can access their data anytime from anywhere [7]. Several open research issues have been identified based on the security and privacy analysis of existing authentication in D2D communication and the weaknesses in some of the recent access control techniques proposed. Even though some of the issues may seem obvious, all these identified open issues could eventually hinder the development and progression of D2D communication. These open research issues would assist future research directions for industries and researchers to secure D2D communication for future practical applications. All these research directions are provided with an emphasis on the security and privacy aspects of D2D communication. The issues identified are listed below:

1) An industry-wide security standard lightweight AKA protocol for secure D2D communication that can incorporate all the different modes of D2D communication mentioned in Section II is currently missing and required. The AKA protocol must accommodate all the three different modes of D2D communication discussed earlier. In addition, the protocol should also be integrated with the traditional cellular network systems.
2) The standalone mode and relay-coverage mode of D2D communication provide a complete and partial decentralisation level, respectively. The government regulatory movements and alliances such as the "Five Eyes" intelligence network [45] and "The Assistance and Access Act of 2018" [46] aim in getting access to the encrypted data that is transmitted over different networks to ensure national and public safety. On the other hand, the standalone mode of D2D communication would provide limited data access to the government. Hence the government is expected to implement stricter regulations in implementing D2D, which can be foreseen as an upcoming issue.
3) Trust and compliance is another issue in the standalone mode of D2D communication. A practical method to authenticate devices in standalone mode and provide fine-grained access control is not discussed in the literature review. Verifying the authenticity of a UE that tries to establish a connection with another UE in standalone mode is difficult, especially when the devices have never communicated previously.
4) Privacy is another open issue of concern. Most of the existing works emphasise the anonymity of users. However, apart from device identity, other physical parameters of the device and actions performed by different users will leave their activity footprints. These footprints could reveal the identity of the user. In addition, they do not want companies to use their personal information for any reasons without consent. Lack of adequate security and privacy evaluation tool can also be a challenge.
5) D2D devices with minimal power consumption and computational capabilities will need to dedicate their security operations resources. Switching between different modes of D2D communication can also be another requirement for these devices. These requirements could add additional overhead to these resource-constrained devices.
6) Non-Repudiation has been identified as a blind spot in most of the existing systems discussed above. It is essential to verify the data, its origination and reception. The devices that send a message to another device should not be allowed to deny the action of sending the message. For example, an active log that could maintain a record of the previous connections established and encrypted messages/hash values can be integrated into future devices.
7) Mode selection overhead is another issue in D2D communication. There exists a gap in research to enable secure communication between UE's that communicate via D2D and the inter-operators within the same mode of D2D communication. Traditional cellular communication networks can switch between different networks based on a relevant CN's availability, but it is not the same for D2D communication. The usability and adaptability of mobile users are affected if their communication performance is affected while movement as they expect a seamless transition between different application scenarios.
8) A D2D communication protocol that can provide various access control privileges to private information of the participating devices can be advantageous. However, all the data from the devices are not to be shared with any device after authentication. Therefore, a complex and secure system to provide granular access control to the data within the device is required.

## VI. Conclusion

D2D communication is a promising technology with various advantages compared to the existing direct communication technologies in the industry and traditional cellular networks. This article has provided a brief on D2D communication and its different modes of operation and other relevant communication technologies. The security and privacy requirements in D2D, followed by the authentication and access control techniques in the existing D2D communication systems, were discussed and further analysed. Even though D2D communication technology is an upcoming technology, there are still many open issues that need to be addressed. It provides an insight into the future research directions for researchers to actively contribute to the practical implementation and seamless deployment of D2D communication.